\documentclass[aps,a4paper,
twocolumn
]{revtex4}

\usepackage{bm}
\usepackage{mathrsfs}
\usepackage{xcolor,color,graphicx,graphics}
\usepackage[all]{xy}
\usepackage{epsfig,subfigure}
\usepackage{latexsym,amssymb,amsmath,amsfonts} 
\usepackage[english]{babel} 
\usepackage[OT1]{fontenc}
\usepackage[latin1]{inputenc}
\usepackage{makeidx}
\usepackage{hyperref}
\usepackage{color,graphicx,graphics,wrapfig,epsf}

\begin{document}
\title
{Wormhole as a possible accelerator of high-energy cosmic-ray particles }
\author{A.A. Kirillov}
\email{ka98@mail.ru}
\affiliation{Bauman Moscow State Technical University, Moscow,  105005, Russian Federation}

\author{E.P. Savelova}
\email{sep$\_$ 22.12.79@inbox.ru}
\affiliation{Bauman Moscow State Technical University, Moscow,  105005, Russian Federation}

\date{
}

\begin{abstract}
We present the simplest topological classification of wormholes and demonstrate that in open Friedmann models
the genus $n\geq 1$ wormholes are stable and do not require the presence of exotic forms of matter, or any
modification of general relativity.
We
show that such wormholes may also possess magnetic fields. It is found that
when the wormhole gets into a galaxy or a surrounding region, it works as
an accelerator of charged particles. If the income of the energy from radiation is small, such a wormhole works simply as a generator of synchrotron radiation.
Estimates show that the threshold energy of such an accelerator may vary from sufficiently modest energies
of the order  of a few Gev, up to enormous energies of the Planckian order and even higher, depending on wormhole parameters.
\end{abstract}


\maketitle

\section{Introduction}

One of the great challenges of modern astrophysics is the origin of the observed high-energy
cosmic-ray particles (HECRs) \cite{rays1,rays}.
There are a  number of high-energy processes in galaxies which may serve as likely candidates
of such extremely energetic particles, e.g., active galactic nuclei, jets in radio galaxies, etc. \cite{Hillas,E18}.
HECRs may also be related to the decay of dark matter particles \cite{Feng10,grib,Khlop14,Khlop15}. Still the nature of their origin is not firmly established.
In the present paper we suggest a new possible mechanism which may allow to accelerate charged particles till extremely high energies. Such a mechanism involves wormholes whose existence is predicted by general relativity and essentially by different extensions of gravity e.g., see \cite{Vis,Sushk,Sushk13,Vagnozzi,Lobo}.

It is expected that relic cosmological wormholes were created from virtual
wormholes on quantum stage, or during the inflationary period of evolution
of the Universe. In general relativity without exotic matter spherically symmetric wormholes are non-static objects
e.g., see \cite{Lobo:2006},  and they presumably collapse very rapidly. However they may be static in various extended theories where the role of exotic matter is played by a modification of general relativity.
It was recently shown that in the Einstein's theory stable quasi-stationary wormholes without exotic matter do exist
in open Friedmann models \cite{KS16}. Such wormholes have a more general structure and have throat sections in the form of tori or more complicated surfaces. General consideration shows that in the flat space static wormholes of such a kind also require the presence of some portion of exotic matter. In the absence of exotic forms of matter such wormholes do evolve. Still it is not clear how rapid is the rate of their evolution.

Some encouraging results were obtained in the limit when one radius of a torus-like throat tends to infinity. Then the geometry acquires cylindrical configuration. 
In particular, in a series of papers \cite{Bron6,Bron7,Bron8,Bron9} static and stationary cylindrical wormhole
solutions were found and it was demonstrated that asymptotically
flat wormhole configurations do not require exotic matter violating the
weak energy condition.

Primordial wormholes may initially capture some electric and magnetic lines of force.
Therefore, from the formal standpoint every entrance into wormhole throat
will look as if it were carrying some electric and magnetic charges, e.g., see exact solutions in \cite{Lobo:2006}.
Of
course the two entrances into the same wormhole have charges of opposite
signs as it should be. In other words, it is more correct here to speak of electric and magnetic poles of different entrances.
 Therefore, in the general case every entrance into such
primordial wormholes can be described by a mass, an angular momentum, an
electric charge, and a magnetic charge as well. During the evolution the
electric charge decays very rapidly, since the electric field transmits
straightforwardly its energy to charged particles in the primordial plasma.
The magnetic part however do not perform the work and retains. Therefore, it
is natural to expect that magnetic fields (magnetic lines captured by
wormholes) survive and may play the role of the seeds which generate
magnetic fields in intergalactic medium. For example, this may explain the
origin of observed magnetic fields in voids \cite{Planck1}.

It turns out that in galaxies such a wormhole may work as an accelerator of
charged particles or simply as a generator of the synchrotron radiation.
Indeed, charged particles captured by magnetic field of a wormhole may move
only along magnetic lines which start from one entrance and end on the
another entrance. Therefore, they are captured by such a wormhole. The
galactic wind produced by a galaxy and the active galactic nuclear push such
particles and speeds them up. Part of the received energy reradiates as the
synchrotron radiation. This acceleration continues till the moment when
particles reach the stationary state (when the received and reradiated
energies become equal), or till the moment when they gain enough energy to
escape the magnetic trap. In general there exists some threshold for the
energy of such particles (the maximal possible value) which is completely
determined by wormhole parameters and the activity of the galaxy. Moreover,
the same wind acts on the magnetic field of the wormhole and may rotate it
to the equilibrium position, when the direction of magnetic lines coincides
with the direction of the wind. This makes the accelerator to work more
efficiently. We may expect that at least part of high-energy cosmic rays may
be produced by such a scheme. This also may explain the recently reported
break in the teraelectronvolt cosmic-ray spectrum of electrons and positrons %
\cite{rays}. Indeed the two branches of such a spectrum correspond simply
to two different such wormhole accelerators, while the position of the brake
corresponds to the threshold energy for the nearest wormhole.

\section{Topological structure of a general genus $n$ wormhole}

According to the Geroch theorem \cite{Geroch} in general relativity the
topological structure of space does not change. Indeed, consider the
Hamiltonian formulation of general relativity. The metric of the space-time
has the form%
\begin{equation}
ds^{2}=N^{2}dt-g_{\alpha \beta }\left( dx^{\alpha }+N^{\alpha }dt\right)
\left( dx^{\beta }+N^{\beta }dt\right) ,
\end{equation}%
where \thinspace $N$ is the lapse function, $N^{\alpha }$ is the shift
vector and $g_{\alpha \beta }$ is the metric of a space-like manifold $S$.
In the Hamiltonian picture functions $N$ and $N^{\alpha }$ have no dynamic
character and they play the role of Lagrange multipliers, while the
Hamiltonian is%
\begin{equation}
H=\int\limits_{S}\left( NC+N^{\alpha }C_{\alpha }\right) d^{3}x,
\end{equation}%
where
\begin{equation}
C=\frac{1}{\sqrt{g}}\left[ \pi ^{\alpha \beta }\pi _{\alpha \beta }-\frac{1}{%
2}\left( \pi _{\alpha }^{\alpha }\right) ^{2}+g\left( -P\right) \right] ,
\end{equation}%
\begin{equation}
C_{\alpha }=-2\nabla _{\beta }\pi _{\alpha }^{\beta }.
\end{equation}%
Here $P$ is the curvature scalar with the 3-dimensional metric $g_{\alpha
\beta }$, $\nabla _{\alpha }$ defines the covariant derivative with the
metric $g_{\alpha \beta }$, and we assume that $S$ is closed space-like
hypersurface (to avoid surface terms in the action). In the presence of
matter fields one has to add respective terms to the Hamiltonian. The
equations of motion have the standard form%
\begin{equation}
\frac{dg_{\alpha \beta }}{dt}=\frac{\delta H}{\delta \pi ^{\alpha \beta }},\
\ \frac{d\pi ^{\alpha \beta }}{dt}=-\frac{\delta H}{\delta g_{\alpha \beta }}%
,
\end{equation}%
while the variation of the action with respect to lapse and shift gives the
constraints ($G_{\alpha }^{0}$ components of the Einstein equations)%
\begin{equation}
C=0,\ \ C_{\alpha }=0.
\end{equation}

Taken the initial values $\left( g_{\alpha \beta },\pi ^{\alpha \beta
}\right) $ on $S$ at some moment of time $t=t_{0}$, the above equations
define in a unique way the subsequent evolution of the dynamic functions $%
\left( g_{\alpha \beta }(t),\pi ^{\alpha \beta }(t)\right) $. This defines
the map of the space-like hypersurfaces $S(t_{0})\rightarrow S(t)$ which is
a diffeomorphism between $S(t_{0})$ and $S(t)$ (differentiable one-to-one map).
Thus in the Hamiltonian
picture we see that the topological structure of the total space-time always
represents the direct product of the topology of the initial hypersurface $S(t_{0})$ and the
time line $t\in R^{1}$. This is roughly constitute the famous Geroch
theorem, which states that in general relativity topology changes do not
occur. We point out that there exist some speculations based on the fact
that choosing different space-like sections of a single maximally extended
space-time we may obtain different topologies of space-like sections.
Some authors try to interpret this as "topology changes".  But the observer
may try different reference frames and easily verify that such a change is
nothing more, than the effect related to a specific frame. Therefore, such
interpretations are not self-consistent from the physical standpoint.

In this manner we see that in general relativity topological structure of
space is determined by onset, as additional initial conditions. In other
words, it is specified by hands. In conclusion of this section we define
topologies which correspond to a general genus $n$ wormhole (e.g., see the
books \cite{Fomenko,Hem,Vis}).

Consider two copies $\partial M_{n}^{\pm }$ of an arbitrary two dimensional
closed surface $\partial M_{n}$ in $S$. From the topological standpoint the
surface $\partial M_{n}$ represents a sphere with $n$ handles on it. The
portion of space $S$ which gets inside such surfaces $x\in M_{n}^{\pm }$ is
removed, while points on respective two copies of the surfaces $\partial
M_{n}^{\pm }$ are glued\footnote{To avoid misunderstanding we point out that here the gluing procedure
concerns the coordinate map only. The common mistake is that such a gluing automatically leads to a thin-shell wormholes. However on this step the metric tensor does not appear on the stage. It appears latter on when we specify particular initial conditions $\left( g_{\alpha \beta }(t_{0}),\pi
^{\alpha \beta }(t_{0})\right) $ on the resulting space. In other words, it is our free choice to specify
the metric which will correspond to a thin-shell or a regular wormhole. The case of the thin-shell requires however the presence of matter fields on the shell.}. This corresponds to the so-called Hegor diagrams %
\cite{Fomenko,Hem}. Thus, we get a wormhole whose throat section is $\partial
M_{n}$ which is sphere with $n$ handles. The number of handles $n$ is called
the genus of two dimensional surface and we keep this name for the wormhole.
Thus we get a wormhole of the genus $n$. The resulting space is $%
S_{n}=S/M_{n}^{\pm }$ and it's structure determines properties of possible
initial conditions $\left( g_{\alpha \beta }(t_{0}),\pi ^{\alpha \beta
}(t_{0})\right) $ which can be specified on $S_{n}$. In particular, since
points on surfaces $\partial M_{n}^{+}$ and $\partial M_{n}^{-}$ are
identical, the admissible functions $\left( g_{\alpha \beta }(t_{0}),\pi
^{\alpha \beta }(t_{0})\right) $ should obey the respective periodic
conditions, namely, they should coincide on $\partial M_{n}^{+}$ and $%
\partial M_{n}^{-}$ and possess the necessary number of derivatives.
The subsequent evolution is completely determined by
the above Hamiltonian equations.

In the case when the distance between surfaces $\partial M_{n}^{+}$ and $%
\partial M_{n}^{-}$ is sufficiently big, the influence of the entrances into
the wormhole on each other can be neglected. In this case we may restrict to
a wormhole which connects two different copies of the space $S$. We may call
them as two-wold wormholes, contrary to the previous case when both
entrances lay in the same space. The two-wold wormholes are more simple for
the investigation and gain more popularity in literature.

We point out that
topology of space was formed (tempered) during quantum stage of the
evolution of our Universe and a priory all kind of wormholes are admissible
as initial conditions. However subsequent evolution is quite different. In
particular, the simplest wormholes of the genus $n=0$ are highly unstable.
To be stable they requite the presence of exotic matter which does not exist
in nature. Genus $n=0$ wormholes may be made stable in modified theories in which exotic matter is replaced
by an appropriate modification. In other words, in the standard Einstein's general relativity such wormholes collapse and are undistinguished from black holes. This means that in spite the widespread popularity of such
solutions (i.e., spherically symmetric wormholes) we should not expect to
find such objects in astrophysics unless the appropriate modification is experimentally established.

\section{Stable relic wormholes}

In general relativity stable relic wormhole cannot have the genus $n=0$. There is an extended
literature on this subject and we may send readers to respective books and
reviews, e.g., \cite{Vis,Lobo}.

In this section we demonstrate that genus $n\geq 1$ wormholes can be made
stable in open Friedmann models, as it was first shown in \cite{KS16}.
The basic idea comes out from very simple qualitative consideration as follows.
Since the section of a wormhole throat $\partial M_{n}$ is closed, its size
is always restricted in space. When we consider the scattering of a thin ray
of particles on such a wormhole it always diverges upon scattering. This
means that the wormhole throat has always a negative curvature. Therefore,
to find the simplest realization of such a topology we should use the open
cosmological model in which space has a constant negative curvature. We
recall that observational bounds allow our Universe to have very small negative
curvature (within the observational errors \cite{Planck1,Planck2}
$ \Omega _k=0.001\pm 0.002
$).
Moreover, even if our Universe is exactly flat, it contains both overdense regions
which possess a positive curvature and underdense regions with a negative curvature.
The latter in some approximation can be viewed as parts of the Lobachevsky space.
Then, as it was
first demonstrated in \cite{KS16} an arbitrary number of genus $n\geq 1$
wormholes can be constructed simply by factorization of the Lobachevsky
space (the space of the open Friedmann model) over a discrete subgroup of
the group of motion of the Lobachevsky space.

We point out that the genus $%
n=0$ wormhole cannot be obtained by such a factorization, since the sphere
does not admit the metric with a negative curvature. When we try to specify
a negative curvature on the sphere, it will contain at least one singular
point. The simplest wormhole (whose existence do not require exotic matter)
has the genus $n=1$ which corresponds to the throat section in the form of a
torus. For details we send readers to our paper mentioned before.

\begin{figure}
\psfig{figure=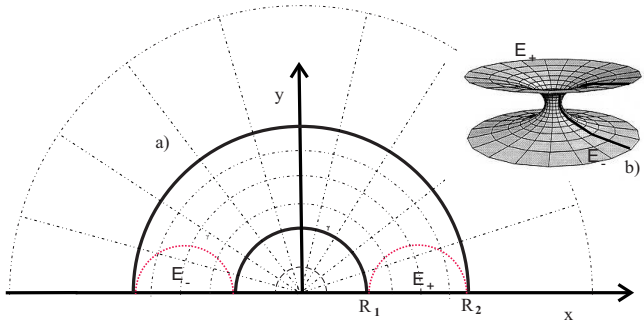,width=8.7cm}
\caption{a)The Lobachevsky plane. Dashed lines give the polar
frame. The stripe between the two geodesic lines $r=R_1$ and
$r=R_2$ corresponds to the wormhole region. Regions below red
dashed geodesics (semi-circles) correspond to two unrestricted
Lobschevsky spaces $E_{\pm}$. b) The same wormhole which connects
two Lobachevsky spaces $E_{\pm}$. Thick solid line corresponds to
the two geodesics $r=R_{1,2}$ on the plane which are glued. } \label{fig1}
\end{figure}

For the illustration of the basic idea we describe the realization of the
simplest two-dimensional wormhole which connects two different spaces by a
factorization of the Lobachevsky plane. Consider the open Friedmann model
which is described by the metric%
\begin{equation}
ds^{2}=c^2dt^{2}-a^{2}(t)\frac{4(dx^{2}+dy^{2}+dz^{2})}{\left( 1-r^{2}\right) ^{2}}.
\end{equation}%
Here the space metric $dl^{2}=\frac{4dx^{2}}{\left( 1-x^{2}\right) ^{2}}$
corresponds to the Lobachevsky space. In 2D case the space can be realized
as a upper complex half -plane Fig\ref{fig1},  when the metric takes the form ($z=x+iy$, $%
y>0 $)
\begin{equation}
dl^{2}=\frac{dx^{2}+dy^{2}}{y^{2}}.  \label{LM}
\end{equation}%
In this metric the line $y=0$ corresponds to the absolute, i.e., the
infinity of the Lobachevsky plane).
Consider now two an arbitrary geodesic lines on the Lobachevsky plane. To
avoid misunderstanding we stress that we are speaking here of space-like
geodesics on the Lobachevsky plane (do not mix them with space-time
geodesics that correspond to motions of probe particles). Recall that
geodesics on the plane are semi-circles with centers on the absolute $y=0$
and perpendicular to the absolute rays ($y=const$). Using the group of
motions of the plane we may choose the coordinate frame on the plane in such
a way that the both geodesics become semi-circles with centers at the origin
$z=0$ but having different radii $R_{1}$ and $R_{2}$. For definiteness we
assume $R_{1}>R_{2}$. Then the geodesic line $y=0$ determines the shortest
distance between these two geodesics. According to (\ref{LM}) we see that
the distance is simply $a=\ln \left( R_{1}/R_{2}\right) $. The simplest
wormhole is obtained by making the cut of the stripe between these two
geodesics and the subsequent gluing of the boundaries (the two geodesics) see Fig \ref{fig1}. The
regions on the plane which are restricted by auxiliary two geodesic lines
with the radius $r=\left( R_{1}-R_{2}\right) /2$ and centers at the
positions $x_{\pm }=\pm (R_{2}+r)$ correspond to two different Lobachevsky
spaces (any geodesic there has infinite length). The rest part of the stripe
has a finite square/volume and corresponds to the throat of the wormhole.

Consider the particular motion $T_{a}$ of the plane that transforms one
geodesic (of the radius $R_{2}$) to the other ($R_{1}$). This is reached by
the transformation
\begin{equation}
z^{\prime }=T_{a}(z)=\frac{R_{1}}{R_{2}}z=e^az.
\end{equation}%
This transformation corresponds to the shift of the plane along the geodesic
line $y=0$ on the distance $a$. One can straightforwardly verify that this
transformation does not change the metric (\ref{LM}) and, therefore, it is
one of particular admissible motions of the plane. Using $T_{a}$ as the
basis element we define the discrete subgroup $G_{a}$ of the group of
motions of the plane which consists of all elements of the form $%
\{T_{a}^{n}\}$, where $n=0,\pm 1,\pm 2,...$. Then the wormhole is merely the
factorization of the plane over $G_{a}$. This means that any two points of
the plane $z$ and $z^{\prime }$ correspond to the same single point of the
wormhole, if they can be related by the transformation $z^{\prime }=\left(
R_{1}/R_{2}\right) ^{n}z=e^{na}z$ with any integer $n$.

The basic difference between the wormhole and the total Lobachevsky plane is
that now we cannot specify an arbitrary functions on the plane. Those are
restricted by a specific "periodic" conditions. For example, if we consider
perturbations of the metric $\delta g_{\alpha \beta }(z)$, then they should
obey the "periodic" property
\begin{equation}
\delta g_{\alpha \beta }(z)=\delta g_{\alpha \beta }(\frac{R_{1}}{R_{2}}z).
\label{Tr}
\end{equation}%
In particular, our attempts to describe dark matter phenomena by wormholes %
\cite{KS07,KS11} owe to this feature. Indeed, if we have a point source $M$
at some position $z_{0}$ (the density is $\varrho =\frac{M}{\sqrt{g}}\delta
(z-z_{0})$), then the condition (\ref{Tr}) produces a countable set of
"additional" sources ($\varrho \rightarrow \varrho =\frac{M}{\sqrt{g}}%
\sum_{n}\delta (z-e^{na}z_{0})$) and this essentially changes the "expected"
standard Newton's potential produced by the source, while the transformation
$\delta (z-z_{0})\rightarrow \sum_{n}\delta (z-e^{na}z_{0})$ we call the
topological bias of sources. This additional images may play the role of
dark matter particles.

We point out that in the case of 3D Lobachevsky space the simplest 3D
wormhole is obtained by analogous factorization over the discrete subgroup
that is formed from two such generators $T_{a}(l_{1})$ and $T_{b}(l_{2})$
which describe two such shifts in orthogonal directions ($l_{1}$ and $l_{2}$
denote two orthogonal geodesics). In this case the minimal section of the
throat will have the form of a torus $T^2$.

Such a factorization can be straightforwardly seen in terms of new
coordinate frame which can be introduced on the Lobachevsky space. Indeed,
consider realization of 3D Lobachevsky space in the form of half space
analogous to (\ref{LM}) which is given by the metric
\begin{equation}
dl^{2}=\frac{dx^{2}+dy^{2}+dz^{2}}{z^{2}}
\end{equation}%
where $z>0$. Consider first the transformation $T_{a}(l_{1})$. Using
possible motions of the space any map $T_{a}(l_{1})$ can be realized as the
conformal transformation $T_{a}\left( r,\phi ,\theta \right) =\left( \frac{%
R_{1}}{R_{2}}r,\phi ,\theta \right) $, where we have used the spherical
coordinate frame $x=r\cos \phi \sin \theta $, $y=r\sin \phi \sin \theta $, $%
z=r\cos \theta $, with variables ranges as $r>0$, $0<\theta <\frac{\pi }{2}$%
, and $0<\phi <2\pi $. In this coordinates the 2D Lobachevsky plane Fig \ref{fig1}.
corresponds to any section $\phi = const$. In this realization the geodesic line $l_{1}$
corresponds simply to the axis $z$ ($l_{1}=(0,0,z)$). Then we may use the
subsequent transformation in the form $r=R_{2}$ $\exp \left( \frac{a}{2\pi }%
\chi _{1}\right) $ which transforms the line element into%
\begin{equation}
dl^{2}= \frac{1}{\cos ^{2}\theta }\left( \left( \frac{a}{2\pi }\right)
^{2}d\chi _{1}^{2}+\sin ^{2}\theta d\phi ^{2}+d\theta ^{2}\right) .
\label{LM3}
\end{equation}%
The region $R_{2}<r<R_{1}$ corresponds to the range of the new angle
variable $0<\chi _{1}<2\pi $, while the total space corresponds to the
unrestricted range of the variable $\chi _{1}$, i.e., $-\infty <\chi
_{1}<+\infty $. In this case the transformation $T_{a}(l_{1})$ acts simply
as the the shift of $\chi _{1}$
\begin{equation}
T_{a}\left( \chi _{1},\phi ,\theta \right) =\left( \chi _{1}+2\pi ,\phi
,\theta \right) .
\end{equation}%
The factorization means that all functions are periodic in the terms of the
new angle variable $\chi _{1}$, i.e., $\delta g_{\alpha \beta }(\chi
_{1})=\delta g_{\alpha \beta }(\chi _{2}+2\pi n)$, with $n=0,\pm 1,\pm 2,...$%
. Absolutely analogously the transformation $T_{b}(l_{2})$ allows to
introduce the second angle-like variable $\chi _{2}$, which gives $%
T_{b}\left( \chi _{2},\widetilde{\phi },\widetilde{\theta }\right) =\left(
\chi _{2}+2\pi ,\widetilde{\phi },\widetilde{\theta }\right) $. Taking now $%
\chi _{1}$ and $\chi _{2}$ as basic coordinates and adding an arbitrary
independent complimentary coordinate variable $\chi _{3}$ we obtain 3D
wormhole as the factorization of the Lobachevsky space in which all points
of the type $\left( \chi _{1}+2\pi n_{1},\chi _{2}+2\pi n_{2},\chi
_{3}\right) $ correspond to the same single point $\left( \chi _{1},\chi
_{2},\chi _{3}\right) $. We point out that in terms of coordinates $\left(
\chi _{1},\chi _{2},\chi _{3}\right) $ the metric (\ref{LM3}) has a rather
complex form and we do not present it here. In particular, the fact that the
factorization does not induce additional curvature terms (and therefore it
does not require exotic matter) can be seen straightforwardly from the
metric (\ref{LM3}).

It is important that such a factorization allows us to obtain an arbitrary
number of wormholes \cite{KS16}. In conclusion of this section we point out
that perturbations in the space which contains such wormholes develop in the
standard way as it is described by the Lifshitz theory \cite{L46}, e.g., see Ref. \cite{Muh}. The
basic difference is that admissible functions obey to the periodic
conditions (\ref{Tr}) (some part of modes is absent). This may put some
restrictions on possible shapes of structures formed. In any case we may
state that some portions of space which may contain the wormhole throats can
form gravitationally bounded systems (galaxies). The evolution of wormholes
in sufficiently dense regions of space requires the further investigations.
However, we may expect that the rate of evolution of such objects is not
crucially differs from that of regions without wormholes and, therefore,
they may survive till present days.

We also point out that in the case of a genus $n\geq 1$ wormhole the
spherical symmetry of such an object can be restored, if we perform
averaging over possible orientations of the throat in space. In some
calculations this allows us to use the simplest spherically symmetric $n=0$
wormhole as a basic "first order" model of the standard (genus $n\geq1$)
wormhole.

\section{Vacuum magnetic fields of wormholes}

Nontrivial topology of space which contains a wormhole allows us to get
additional nontrivial solutions of basic physical field equations, equations
of motion, and, in particular, of static vacuum Maxwell equations. This
means that already in the absence of real sources (charged particles,
electric currents) space may possess nontrivial quasi-static magnetic and
electric fields. The physical mechanism is rather clear, during the quantum
period of the Universe when the wormhole forms, it may capture some portion
of closed magnetic or electric lines. Such lines cannot simply leave the
wormhole. Electric fields perform work and, therefore, they decay very
rapidly, since in the primordial plasma they transform their energy to
charged particles. However magnetic fields may survive till the present days.
For exact solutions which involve  magnetic field and wormholes see e.g., \cite{Lobo:2006,Bron18}.

In this section for the sake of simplicity we assume that sufficiently far from the wormhole entrances the space is flat. Generalization to the curved spacetime and consideration of
magnetic fields in the Friedmann model can be found, e.g., in \cite{B16}.
Consider first the simplest genus $n=0$ wormhole. Then the space-time metric can
be taken as
\begin{equation}
dl^{2}=c^2dt^2-h^{2}(r)\left( dx^{2}+dy^{2}+dz^{2}\right) .  \label{whm}
\end{equation}%
We shall use the  Ellis--Bronnikov massless wormhole \cite{Bron,Ellis} when the scale function is simply $h=1+\frac{R^{2}}{r^{2}}$.  In what follows for simplicity we may replace this function with the model
\begin{equation}
h(r)=\left\{
\begin{array}{c}
1,\ r>R \\
\frac{R^{2}}{r^{2}},\ r<R%
\end{array}%
\right. .
\end{equation}%
This thin--shell model of a wormhole is described by two Euclidean spaces $E_{+}$ as $r>R$ and $%
E_{-}$ as $r<R$. The transformation $\widetilde{r}=R^{2}/r$ does not change
the above metric but interchanges the spaces $E_{+}\leftrightarrow E_{-}$.
Both spaces are glued by the surface of the sphere $r=R$. Since the space is
flat (only on the throat at $r=R$ it has delta-like curvature scalar $\sim
\frac{-1}{R^{2}}\delta (r-R)$), the Maxwell equations for magnetic field
take the standard form%
\begin{equation}
rot\mathbf{B}=div\mathbf{B}=0.  \label{Max}
\end{equation}%
These equations possess a nontrivial solution in the form
\begin{equation}
\mathbf{B}=-\frac{Q}{r^{2}h}\mathbf{n}
\end{equation}%
with an arbitrary constant value $Q$, where $\mathbf{n}=\mathbf{r}/r$ is the
unit vector.
Indeed, vacuum magnetic field can be expressed via the magnetic scalar
potential $\mathbf{B}=-\mathbf{\nabla }\phi $ which obeys the equation%
\begin{equation}
\Delta \phi =0.
\end{equation}%
In the spherically symmetric case this equation reduces to%
\begin{equation}
\frac{1}{r^{2}h^{3}}\partial _{r}r^{2}h\partial _{r}\phi =0
\end{equation}%
and has the solution as%
\begin{equation}
\phi =\int_{0}^{r}\frac{Q}{r^{2}h}dr+\phi _{0}.
\end{equation}%
This defines the magnetic field
\begin{equation}
\mathbf{B}=-\mathbf{\nabla }\phi =-\frac{Q}{r^{2}h}\mathbf{\nabla }r=-%
\frac{Q}{r^{2}h}\mathbf{n.}
\end{equation}
This solution works for any spherically symmetric wormhole with an arbitrary
scale function $h(r)$ in (\ref{whm}).

In the region $E_{+}$ ($r>R$) it describes the field of the
magnetic charge $Q$ homogeneously distributed over the sphere $r=R$ (Coulomb
law). In the region $E_{-}$ ($r<R$) transformation $\widetilde{r}=R^{2}/r$
interchanges the inner and outer regions of the sphere $r=R$ and we get the
same field with the magnetic charge $-Q$.

In the case when both entrances are in the same space this transforms to the
dipole field. Let the positions of the two spheres are $\mathbf{x}_{+}$ and $%
\mathbf{x}_{-}$ then the field can be taken as ($h_{\pm}=h(r_{\pm})\sim 1$ as $r_{\pm}=\left\vert
\mathbf{x}-\mathbf{x}_{\pm}\right\vert\gg R$)%
\begin{equation}
\mathbf{B(x)}=\frac{Q\left( \mathbf{x}-\mathbf{x}_{+}\right) }{h_+\left\vert
\mathbf{x}-\mathbf{x}_{+}\right\vert ^{3}}-\frac{Q\left( \mathbf{x}-\mathbf{x%
}_{-}\right) }{h_-\left\vert \mathbf{x}-\mathbf{x}_{-}\right\vert ^{3}}.
\label{DF}
\end{equation}%

Consider now the genus $n=1$ wormhole. In the flat space such a wormhole can
be constructed by means of cutting two solid tori and gluing along their
surfaces. In this case we have two different new kinds of solutions. First
is obtained by placing an arbitrary magnetic charge density $\rho (x)$ in
the internal region of one torus and the opposite density $-\rho (x)$ in the
internal region of the second torus. Then we solve the system
\begin{equation}
rot\mathbf{B}=0,\ \ div\mathbf{B}=4\pi \rho .  \label{1a}
\end{equation}%
Recall that internal regions of tori correspond to fictitious points, while $%
\rho (x)=0$ as $x$ lies outside the tori. Therefore, such a system coincides
exactly with (\ref{Max}). In this case the exact form of the magnetic field
is rather complicated. However averaging over orientations of the tori we
restore the spherical symmetry of the wormhole and get exactly the same
solution as (\ref{DF}). In this sense for the sake of simplicity we may
always restrict to the spherically symmetric wormholes (i.e., consider $n=0$
wormhole as the first order approximation).

Solutions of the second kind can be obtained by placing an arbitrary current
density $\mathbf{j}(x)$ and $-\mathbf{j}(x)$ within the two tori and solving
the system%
\begin{equation}
rot\mathbf{B}=\frac{4\pi }{c}\mathbf{j},\ \ div\mathbf{B}=0.  \label{1b}
\end{equation}%
Again here the non-vanishing current density corresponds only to fictitious
points, while in physical regions of space such a system represents the same
system (\ref{Max}). The system (\ref{1b}) corresponds to the field generated
by a couple of electric loops. We point also out that solutions of this
system cannot be reduced to the spherically symmetric case.

In conclusion of this section we point out that the new two classes of vacuum solutions
reflect the topological non-triviality of space. Indeed, according to the Stocks theorem
the system (\ref{Max}) implies $\oint \mathbf{Bdl}=0$ for any loop  which can be pulled to a point.
In the case of a non-trivial topology of space there appear new classes of loops $\Gamma_a$ which cannot be contracted to a point and, therefore, to fix the unique solution we have to fix additional boundary data $\oint _{\Gamma_a}\mathbf{Bdl}=C_a$. In general $C_a\neq 0$. In the case of genus $n=0$ wormhole there is only one such a non-trivial loop which goes through the wormhole throat. In the case of genus $n=1$ wormhole we have already two such loops, one goes through the throat (it corresponds to the system (\ref{1a})) and one additional crosses the torus (the system (\ref{1b})).

\section{wormhole as an accelerator}

\begin{figure}[tbp]
\centerline{\psfig{figure=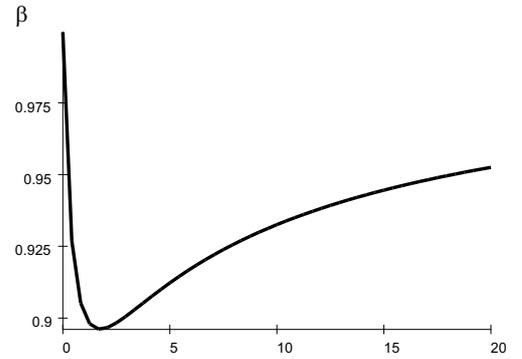,width=6.5cm}}
\caption{ The spectral coefficient $\beta
_{\nu }$ as the function of $x =\left( h\nu +cp\right) \left(
E+cp\right) /m^{2}c^{4}$.} \label{fig2}
\end{figure}

In galaxies charged particles undergo an acceleration when interacting with
galactic radiation. For definiteness we shall speak of electrons. Indeed, by
means of the Compton scattering photons transmit part of their momentum to
electrons. Upon the scattering the momentum obtained by the electron from
the incident photon is

\begin{equation}
\mathbf{\Delta p}=\mathbf{p^{\prime }-p}=\frac{h\nu}{c}\left( \mathbf{n}-\frac{\nu ^{\prime }}{\nu }\mathbf{n}^{\prime
}\right)
\end{equation}%
where $\mathbf{n}$ and $\mathbf{n}^{\prime }$ are the direction of the
photon before and after scattering and
\begin{equation}\label{nu}
\frac{\nu ^{\prime }}{\nu }=\frac{\left(1-\frac{c}{E}\mathbf{p n
}\right)}{\left( 1+\frac{h\nu }{E}\left( 1-\mathbf{nn
^{\prime }}\right) -\frac{c}{E}\mathbf{p n
^{\prime }}\right)}.
\end{equation}%
Upon averaging over possible orientations of $\mathbf{n}^{\prime }$ we get
for the average momentum transmitted from the incident photon
\begin{equation}\label{pi}
\mathbf{\Delta p}=\frac{1}{c}\beta _{\nu }(\mathbf{p}) h\nu \mathbf{n}.
\end{equation}%
Here the spectral coefficient $\beta _{\nu }(\mathbf{p})=1-\left\langle \frac{\nu
^{\prime }}{\nu }\mathbf{nn}^{\prime }\right\rangle $ is given by
\begin{equation}
\beta _{\nu }(\mathbf{p})=1-\frac{1}{4\pi }\int \left( \frac{\nu ^{\prime }}{\nu }%
\mathbf{nn}^{\prime }\right) d\Omega ^{\prime },
\end{equation}
where $\frac{\nu ^{\prime }}{\nu }$ is determined by (\ref{nu}).
In the case when $\mathbf{p}=p\mathbf{n}$, it reduces to the form
\begin{equation*}
\beta _{\nu }=1+\frac{1}{2x} \left[2-\left( 1+\frac{1}{x}\right) \ln \left( 1+2x \right) \right],
\end{equation*}
where $x =\left( h\nu +cp\right) \left( E+cp\right) /m^{2}c^{4}$.
We plot this function on Fig\ref{fig2}.
Now multiplying (\ref{pi}) on the number density of photons with the frequency $%
\nu $ and on the cross section we get the spectral force which accelerates
the electron in the form%
\begin{equation}
\mathbf{f}_{\nu }=\frac{\Delta \mathbf{p}}{\Delta t}=c\sigma _{T}N_{\nu }%
\frac{1}{c}\beta _{\nu }(p) h\nu \mathbf{n=}\frac{\sigma _{T}\beta _{\nu }(p)
}{c} \mathbf{P}_{\nu }
\end{equation}%
where $\mathbf{P}_{\nu }$ is the spectral component of the Poynting vector $%
\mathbf{P_{\nu }=}\frac{c}{4\pi }\mathbf{E_{\nu }\times B}_{\nu }$ and $%
\sigma _{T}$ is the Thomson cross section. The total force is given by $%
\mathbf{F=}\int \mathbf{f}_{\nu }d\nu $.

It is important that the force is determined by the Poynting's vector $%
\mathbf{P}_{\nu }$. In a quite (quasi-stationary or steady state) galaxy
both the Poynting's vector and the force have the potential character, i.e.,
they can be presented as $\mathbf{P_{\nu }=-\nabla }\Psi _{\nu }$.
Non-stationary processes in active galactic nuclei may produce some
additional acceleration, e.g., the stochastic Fermi acceleration, etc.,
which we do not discus here. For the quasi-stationary galaxy the Poynting's
theorem gives the discontinuity equation
\begin{equation}  \label{PT1}
div\mathbf{P_{\nu }=\ell }_{\nu }
\end{equation}%
where $\ell _{\nu }(x)$ is the spectral density of sources of radiation
(stars, hot gas, dust, etc.) or the radiative capability of a unite volume
in the galaxy. If the topology is simple, then sufficiently far from the
galaxy we get%
\begin{equation}
\mathbf{P_{\nu }=}\frac{M_{\nu }}{4\pi r^{2}}\mathbf{l,}
\end{equation}%
where $\mathbf{l=r/}r$, $\mathbf{r}$ is the distance from the center of the
galaxy, and $M_{\nu }$ is the total spectral energy emitted by the galaxy in
the unit time. Observations show that the intergalactic medium possesses a
magnetic field \cite{Planck1,B2}. Therefore, the electron may have a closed
trajectory. It is easy to verify that the total energy obtained by the
electron from the galactic radiation during the cycle is exactly zero, i.e.,
$\oint \mathbf{f_{\nu }dl}\equiv 0$. This means that in the case when
topology is simple, the only possible mechanism of the electron acceleration
relates to high-energy non-stationary processes (jets, shock waves, supernovae explosions,
active galactic nuclei, etc.).

The situation changes  when the galaxy is accompanied with a
wormhole. In the presence of the wormhole the Poynting's field also admits
non-trivial solutions of (\ref{PT1}). For the sake of simplicity we consider
the spherically symmetric (genus $n=0$) wormhole. For a more general wormhole the rough picture remains the same, at least from the qualitative standpoint. We point out that the genus $n\geq 1$ wormholes are more
preferred from the astrophysical standpoint, since they may work
as accelerators even in the case when a wormhole is not traversable (e.g., when the length of the trajectories which
go through the throat are too big).

Indeed, the scattering of the radiation
on the wormhole (e.g., see \cite{KS18,KS19}) produces an additional field
in the dipole form%
\begin{equation}
\delta \mathbf{P_{\nu }=}\frac{\delta M_{\nu }}{4\pi r_{-}^{2}}\mathbf{%
\mathbf{n}_{-}-}\frac{\delta M_{\nu }}{4\pi r_{+}^{2}}\mathbf{n}_{+}
\label{PV}
\end{equation}%
where $\mathbf{n}_{\pm }=\mathbf{r}_{\pm }/r_{\pm }$, $\mathbf{r}_{\pm }=\mathbf{r}-\mathbf{x}_{\pm }$, $\mathbf{x}_{\pm }$
are positions of the wormhole entrances (we assume that $r_{+}\ll r_{-}$), $%
\delta M_{\nu }\simeq \frac{M_{\nu }\pi R^{2}}{4\pi r_{+}^{2}}$ is the
portion of the spectral energy absorbed by the closest entrance into the
wormhole throat and $R$ is the radius of the throat. If the wormhole possesses a
magnetic field in the form (\ref{DF}), it forms a magnetic trap for the
electron, while the Poynting's vector field $\delta \mathbf{P_{\nu }} $ in
form (\ref{PV}) forms the accelerating force which acts exactly along the
magnetic lines. On every cycle the electron will gain the energy from
radiation $A=\int A_{\nu }d\nu >0$, where $A_{\nu }=\oint \delta \mathbf{%
f_{\nu }dl}$, whose exact value depends on the length of the trajectory of
the electron and on all the rest parameters of the wormhole (distance to the
galaxy, throat size, etc.). Some part of this energy will be spent on the
synchrotron radiation of electrons and, therefore, there is a competition
between the acceleration produced by the galactic radiation and loss of
energy on the reradiation. The reradiation can be accounted for by adding the
standard force of the radiation friction.

\section{Conclusions}

In this manner we see that the system which consists of a galaxy and a
wormhole endowed with a magnetic field works as an eternal accelerator.
While particles have sufficiently small energy they are trapped and the
system accelerates them. Particles of sufficiently high energies may leave
the trap and contribute to the observed high-energy cosmic rays. The maximal
possible energy which can be reached in such an accelerator depends on two
basic factors.

The first factor is that due to the curvature of the magnetic
lines electrons always have the component of the velocity which is
orthogonal to the magnetic field. This causes the synchrotron radiation
which gives the loss of the energy. For sufficiently high energies such a
radiation becomes very strong and this restricts the maximal possible energy
reached by such an accelerator (the exact value depends on all parameters of
the accelerator). If the equilibrium between the acceleration and the loss of energy
is reached, the wormhole starts to work simply as a generator of synchrotron
radiation. Such objects should be directly seen on sky (the map of such sources
is presented by \cite{Planck1}). The problem of relating such sources to wormholes
requires further and more rigorous investigation.

The second factor appears
when such a limit is not reached (i.e., when the loss of the energy still does
not exceed the obtained from the radiation value), then the absolute
estimate of the threshold energy can be found as follows. The magnetic field
reaches the maximal value at the throat and it is given by
\begin{equation}
B_{w}=\frac{Q}{R^{2}}.
\end{equation}%
Now consider the relation
\begin{equation}
V_{\perp }=r_{B}\omega _{B}=r_{B}\frac{eQ\gamma }{mcR^{2}}=r_{B}\frac{eB_{w}c}{E},
\end{equation}%
where $V_{\perp }$ is the component of the velocity transversal to the
magnetic field (one may assume that $V_{\perp }\sim V_{\Vert } \sim c$), $\omega
_{B}$ and $r_{B}$ are the Larmor's frequency and radius respectively, $%
\gamma =(1-V^{2}/c^{2})^{1/2}$ is the relativistic factor, $e$ and $m$ are
the electron charge and mass respectively, and $R$ is the size of the
wormhole throat. To undergo the acceleration the electron should always get
into the wormhole throat. This means that the Larmor's radius should be
smaller than the throat radius $r_{B}=c/(\omega_{B})=(cE)/(eB_{w}c)\leq R$.
For ultra-relativistic
particles this inequality determines the absolute threshold energy as
$$
E\leq E_{th}=eB_{w}R.
$$
Particles of higher energies cannot get into the throat and stop receiving
the energy from the radiation.  To find estimates we point out that the value $B_{w}$ can be related to the observed value in galaxies and clusters by the relation
$$B_{obs}\sim B_{w}(\frac{R}{L})^2,
$$
where $R$ is the radius of the throat and $L$ is the distance between the wormhole entrances.
Then for the threshold energy we get the estimate energy as
$$E_{th}=eLB_{obs}\frac{L}{R}.
$$
Now taking $R_{\bigodot}$ be the Sun radius, $L=L_{5}\times 5kpc$ (in dark matter models based on wormholes this corresponds to the scale when
dark matter starts to show up \cite{KS07,KS11}), and $B_{obs}=B_{-9}nG=B_{-9}10^{-9}$ Gauss, we get the estimate
$$E_{th}\sim L^2_5B_{-9}\frac{R_{\bigodot}}{R}\times10^{17}Gev.
$$
We see that from the formal standpoint the threshold may reach and even exceed the Planckian value $\sim 10^{19}Gev$. However, for sufficiently small values of the throat radius we may expect that the matter density in the throat is very high. Therefore, the scattering on baryons in the throat will close such an accelerator for particles. In this sense the throat will be not traversable for charged particles.
Consider now the estimate for torus-like (genus $n=1$) wormhole. In this case we may expect the value $R\sim 10^{-3}pc$ which gives the factor $R_{\bigodot}/R \sim 2.3 \times 10^{-15}$ and the threshold energy estimate becomes
$$E_{th}\sim 2 \L_{5}B_{-9}\times 100Gev .$$
This is already close to the reported in \cite{rays} value for the break energy $ E_{b}=(9.14\pm 0.98)100$Gev.

The intensity of the flow of high-energy
particles leaving such an accelerator is determined by the luminosity of the
galaxy, the distance between the wormhole and the galactic center, and the
cross-section of the wormhole which has the order $\sim \pi R^{2}$.

In conclusion  we point out that if we have two such
wormholes near the same galaxy (or even in two different galaxies), then
they should produce different contributions to the high-energy cosmic-ray
spectrum depending on parameters of the wormholes, activity of the galaxy,
etc. This allows us to interpret the break of the high-energy spectrum observed in \cite{rays} as
contributions from different wormholes. In general wormholes have different
threshold energies. The break indicates that at least one of wormholes has
the threshold energy $E_{th}\simeq E_{b}$. This
however still does not allow us to derive rigorous estimates to all wormhole
parameters $L$, $R$, and $B_w$.

\end{document}